\documentclass[aps,amsmath,prapplied,reprint,groupedaddress]{revtex4-1}
\usepackage{graphicx}
\usepackage{dcolumn}
\usepackage{bm}
\usepackage{amsfonts}
\usepackage{array}
\usepackage{booktabs}
\usepackage{makecell}
\usepackage{threeparttable}
\usepackage[colorlinks,linkcolor=red,anchorcolor=blue,citecolor=green]{hyperref}

\begin{document}


\title{Rate compatible reconciliation for continuous-variable quantum key distribution using Raptor-like LDPC codes}


\author{Chao~Zhou$^{1}$}
\author{Xiangyu Wang$^1$}
\email[]{xywang@bupt.edu.cn}
\author{Zhiguo Zhang$^1$}
\author{Song Yu$^1$}
\author{ZiYang Chen$^2$}
\email[]{chenziyang@pku.edu.cn}
\author{Hong Guo$^2$}

\affiliation{$^1$State Key Laboratory of Information Photonics and Optical Communications, Beijing University of Posts and Telecommunications, Beijing 100876, China}
\affiliation{$^2$State Key Laboratory of Advanced Optical Communication, Systems and Networks, Department of Electronics, and Center for Quantum Information Technology, Peking University, Beijing 100871, China}


\date{\today}

\begin{abstract}

In the practical continuous-variable quantum key distribution (CV-QKD) system, the postprocessing process, particularly the error correction part, significantly impacts the system performance. Multi-edge type low-density parity-check (MET-LDPC) codes are suitable for CV-QKD systems because of their Shannon-limit-approaching performance at a low signal-to-noise ratio (SNR). However, the process of designing a low-rate MET-LDPC code with good performance is extremely complicated. Thus, we introduce Raptor-like LDPC (RL-LDPC) codes into the CV-QKD system, exhibiting both the rate compatible property of the Raptor code and capacity-approaching performance of MET-LDPC codes. Moreover, this technique can significantly reduce the cost of constructing a new matrix.
We design the RL-LDPC matrix with a code rate of 0.02 and easily and effectively adjust this rate from 0.016 to 0.034.
Simulation results show that we can achieve more than 98\% reconciliation efficiency in a range of code rate variation using only one RL-LDPC code that can support high-speed decoding with an SNR less than -16.45 dB.
This code allows the system to maintain a high key extraction rate under various SNRs, paving the way for practical applications of CV-QKD systems with different transmission distances.

\end{abstract}


\maketitle

\section{Introduction}
Quantum key distribution (QKD) \cite{pirandola2019advances,RevModPhys.92.025002,diamanti2016practical} enables two spatially separated parties—named Alice and Bob—for extracting a symmetrical string of secret keys using the fundamental properties of quantum mechanics. Currently, there are two mainstream schemes for QKD research, namely, discrete-variable QKD (DV-QKD)\cite{bennett2014quantum,gisin2002quantum,scarani2009security} and continuous-variable QKD (CV-QKD) \cite{grosshans2002continuous,grosshans2003quantum,braunstein2005quantum,weedbrook2012gaussian}. In DV-QKD systems, the polarization or phase of a single-photon state is encoded by key information, whereas in CV-QKD systems, the amplitude and phase quadrature of quantum states are encoded. The CV-QKD system offers good application prospects for the implementation of classical telecom components, thus attracting considerable research attention. Research activities have primarily focused on extending the transmission distance and improving secret key rate between two parties in the CV-QKD systems\cite{jouguet2013experimental,Zhang_2019,zhang2020PRL,GUO2021}.

A typical CV-QKD system consists of two phases. First, Alice prepares a quantum state and transmits it to Bob through a quantum channel; Bob uses a homodyne or heterodyne detector to measure the raw data. Second, both parties extract the secret key by performing postprocessing through a classical channel. In postprocessing, the information reconciliation step, particularly the performance of error correction, will limit the transmission distance and secret key rate of the system. Currently, commonly used reconciliation methods are slice reconciliation \cite{PhysRevA.90.042329} and multidimensional reconciliation \cite{leverrier2008multidimensional,jouguet2011long}. The slice reconciliation method is more suitable for short-distance transmission systems and can extract multiple bits of information from a single pulse. The multidimensional reconciliation method provides an effective way to correct low signal-to-noise ratio (SNR) Gaussian continuous variables using classical error correction codes \cite{zhaoshenmei,baizengliang}. Increased reconciliation efficiency, which can be achieved using near-Shannon-limit error correction codes with low rates in the information reconciliation step, is a technical challenge in extending transmission distance and improving secret key rate of CV-QKD systems \cite{milicevic2018quasi,zhou2019}.

Multi-edge type low-density parity-check (MET-LDPC) codes \cite{richardson2002multi} exhibiting low rates combined with the reverse multidimensional reconciliation scheme can achieve excellent correction performance in the CV-QKD system. The SNR of an optical quantum channel is low in such a long-distance transmission, thus requiring a low code rate and long code block length. For example, when the SNR is less than -15 dB, the code rate is less than 0.02 and the block length has an order of $10^{6}$ \cite{leverrier2009unconditional}. However, designing a parity-check matrix with good performance is complex, and it is extremely complicated to design all matrices for different SNRs \cite{wang2017efficient}. Therefore, to improve the reconciliation efficiency, the system must change the modulation variance at Alice’s side to ensure the receiving variables achieve the target SNR. The frame error rate and postprocessing speed must also be considered in the practical application of the CV-QKD system \cite{9057665,li2020high}. Multiple interactions and decoding steps of Alice and Bob in the information reconciliation step will increase system delay. This paper focuses on a method achieving good error correction performance that can be flexibly implemented in different scenarios in the practical system.

Further, we introduce Raptor-like LDPC (RL-LDPC) codes \cite{6744572} into the CV-QKD system. These codes possess a rateless Raptor code property as well as good error correction performance of MET-LDPC codes under a low SNR. The raptor-like structure is important for providing rate compatibility and obtaining capacity-approaching performance, particularly at low code rates \cite{7045568,LIU20151582}. The generator matrix of the Raptor codes is randomly generated; thus, it is impossible to remove all short cycles in the matrix, and the error correction performance is unstable. Unlike the Raptor codes \cite{1638543,zhou2019}, the structure of each additional parity bit is explicitly designed through density evolution in the RL-LDPC codes. Currently, the RL-LDPC code is mainly used in wireless broadcast communications. We simply design a RL-LDPC code with a rate of 0.02 and adjust this code rate. Numerical results show good performance of the RL-LDPC codes, which are designed to meet the requirements of the practical CV-QKD system in real-time and realize high-speed processing in different application scenarios.

The remaining of this paper is organized as follows: In Sec.~\ref{sec2}, the structure of RL-LDPC codes is briefly introduced and its application to the postprocessing of the CV-QKD system is provided. In Sec.~\ref{sec3}, we describe the construction of low-rate RL-LDPC codes and realize rate compatibility. In Sec.~\ref{sec4}, several simulations are performed to show the good performance of the RL-LDPC codes. Finally, the conclusions of this paper are summarized in Sec.~\ref{sec5}.

\section{Postprocessing using RL-LDPC codes}
In this section, we introduce the RL-LDPC codes into the CV-QKD system. First, we provide additional details about the RL-LDPC codes. As the name implies, such codes are special LDPC codes with the structure of the Raptor code. Raptor codes are the first rateless codes with linear time encoding and decoding, and they have been used in several applications with large data transmissions. The RL-LDPC codes retain the characteristics of Raptor codes to some extent.

\label{sec2}
\vspace{-0.2cm}
\subsection{Base matrix structure of RL-LDPC codes}
\begin{figure}[]
	\centering
	\includegraphics[width=0.9\linewidth]{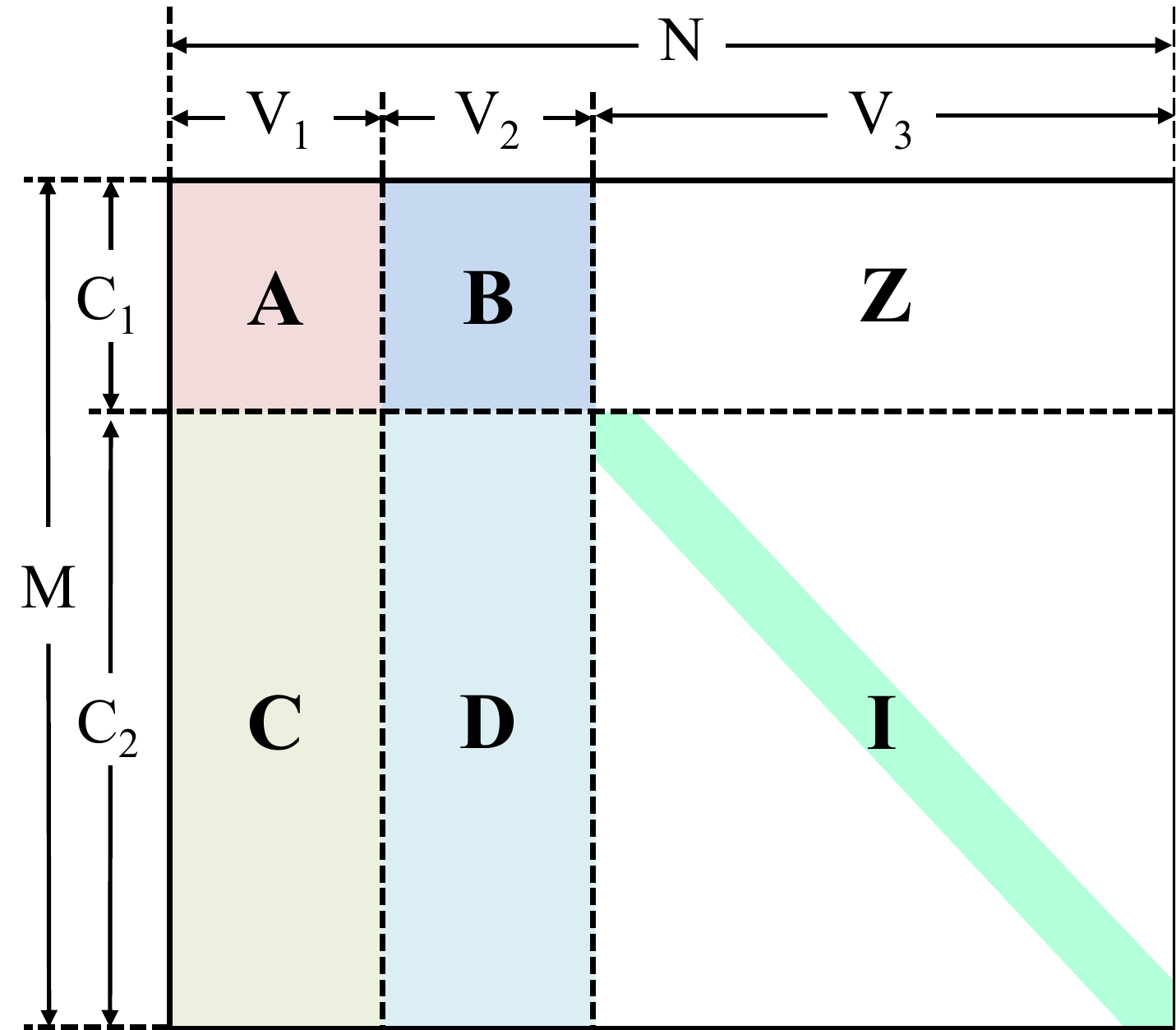}
	\caption{Base parity-check matrix structure of RL-LDPC codes. The size of this base matrix is $M\times N$. $\mathbf{A}$, $\mathbf{B}$, $\mathbf{C}$, and $\mathbf{D}$ are four sparse subparity-check matrices, and their sizes are $C_{1}\times V_{1}$, $C_{1}\times V_{2}$, $C_{2}\times V_{1}$, and $C_{2}\times C_{2}$, respectively. $\mathbf{Z}$ is a $C_{1}\times V_{3}$ zero matrix, and $\mathbf{I}$ is a $C_{2}\times V_{3}$ identity matrix. In practice, the size of the submatrices will be adjusted based on the degree distribution of the MET-LDPC codes.}
	\label{fig:raptor-code}
\end{figure}
\begin{figure*}[t]
	\centering
	\includegraphics[width=1.0\linewidth]{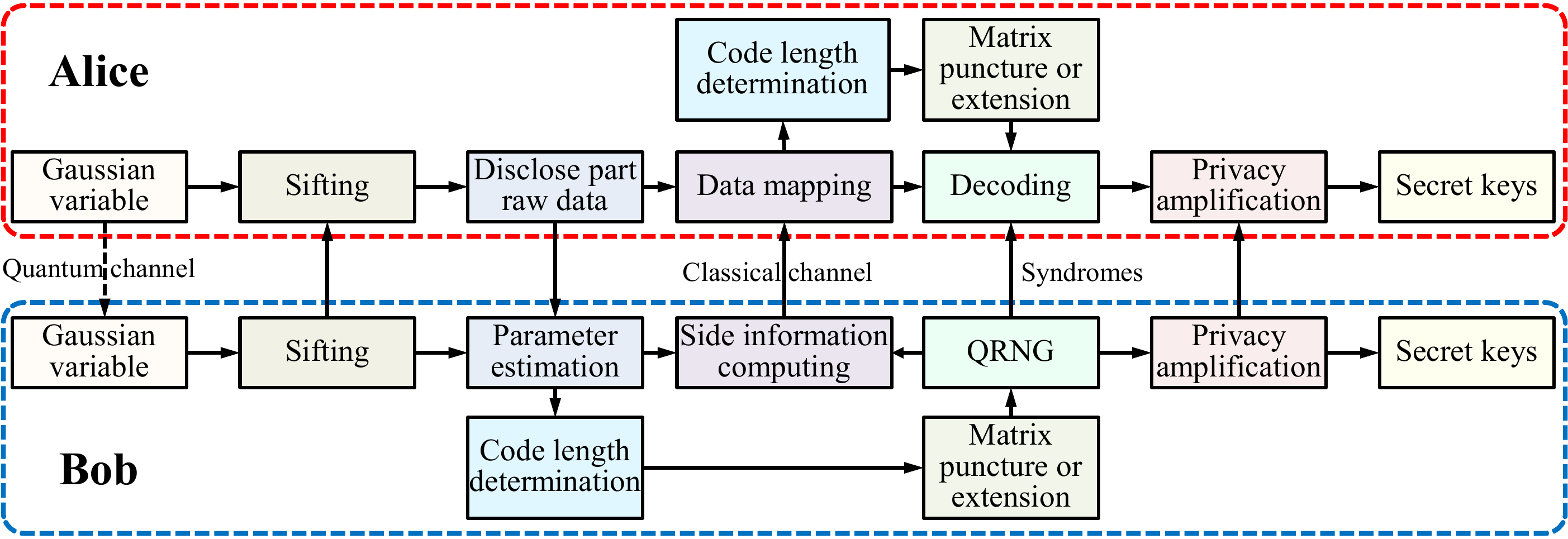}
	\caption{Schematic of postprocessing with reverse reconciliation in the CV-QKD system using RL-LDPC codes. Alice transmits quantum states to Bob through a private optical channel, and both parties extract the secret keys through an authenticated classical public channel. QRNG: quantum random number generator.}
	\label{fig:postprocessing}
\end{figure*}

Generally, the Raptor codes include two parts: linear precoder (such as LDPC codes) and Luby transform (LT) codes. Its Tanner graph is concatenated with LDPC and LT code Tanner graphs and can be transformed into a parity-check matrix, similar to the structure in Figure~\ref{fig:raptor-code}. This figure shows the base parity-check matrix structure of the RL-LDPC codes. As we can see that the base parity-check matrix with a size of $M\times N$ is constructed using different size submatrices, different colors represent different types of edges. $\mathbf{A}$, $\mathbf{B}$, $\mathbf{C}$, and $\mathbf{D}$ are four sparse subparity-check matrices, and their sizes are $C_{1}\times V_{1}$, $C_{1}\times V_{2}$, $C_{2}\times V_{1}$, and $C_{2}\times C_{2}$, respectively. $\mathbf{Z}$ is a zero matrix with a size of $C_{1}\times V_{3}$, and $\mathbf{I}$ is an identity matrix with a size of $C_{2}\times C_{3}$. It can be seen from this figure that the sizes of matrices $\mathbf{A}$, $\mathbf{B}$, $\mathbf{C}$, and $\mathbf{D}$ are related. Hence, their sizes can be adjusted based on their degree distributions, or the two submatrices can be merged into one. If submatrices $\mathbf{A}$ and $\mathbf{B}$ are considered as precoding matrices and $\mathbf{C}$ and $\mathbf{D}$ as generator matrices of LT codes, the structure shown in Figure~\ref{fig:raptor-code} will represent the parity-check matrix of Raptor codes. The RL-LDPC codes can be regarded as special MET-LDPC codes, which consist of different types of edges.

In this paper, we design low-rate LDPC codes based on the long-distance CV-QKD system structure. This structure not only maintains the capacity-approaching performance of MET-LDPC codes but also exhibits the rate compatibility of Raptor codes.

\subsection{Procedure of postprocessing}

Here, we introduce RL-LDPC codes into the CV-QKD system and explain its operation in postprocessing. The schematic of postprocessing is shown in Figure~\ref{fig:postprocessing}. First, Alice prepares Gaussian-modulated coherent states and transmits them to Bob, who measures one of the quadratures using a homodyne detector. After Bob measures the quantum states received from Alice through a quantum channel, both parties start extracting the secret keys through a public channel, which is assumed to be noiseless and error-free.

Postprocessing mainly includes four steps, namely, base sifting, parameter estimation \cite{Leverrier2010Finite,jouguet2012analysis}, information reconciliation \cite{grosshans2003virtual,van2004reconciliation}, and privacy amplification \cite{bennett1995generalized}. Base sifting implies that Bob transmits his selected measurement base to Alice, and then Alice retains the relevant data. Next, Bob determines quantum channel parameters and estimates the secret key rate using the parameter estimation step. The optimal code rate can be calculated using the current SNR and target reconciliation efficiency and can be used to determine the size of the RL-LDPC parity-check matrix needed in the information reconciliation step. According to the optimal code rate, Alice and Bob puncture or extend the base parity-check matrix. Owing to a long transmission distance in this work, we adopt the reverse multidimensional reconciliation scheme. Here, Bob uses a quantum random number generator to generate the initial secret keys to calculate the side information to be transmitted to Alice. Alice determines the size of the matrix of the RL-LDPC code based on the length of the received-side information and then corrects the errors after data mapping. Both parties can hold a set of common sequences when Alice successfully decodes the information. Following the above steps, Eve may have collected sufficient information during her observations of the quantum and classical channels. Hence, privacy amplification is an indispensable step to extract final secret keys from common sequences between Alice and Bob.

Considering the finite-size effects, the secret key rate of the CV-QKD system using one-way reverse reconciliation is expressed as \cite{Leverrier2010Finite,pirandola2019advances,zhang2020PRL}
\begin{equation}\label{finite_secret_key}
K_{finite}=\alpha(1-\mathrm{FER})\left [ \beta I_{AB}-\chi_{BE}-\bigtriangleup(n)\right ],
\end{equation}
where $\alpha$ is the proportion of the number of key extraction in the total number of data exchanged by Alice and Bob,
$\beta \in {[0,1]}$ is the reconciliation efficiency, and $I_{AB}$ is the classical mutual information shared between Alice and Bob. $\mathrm{FER}$ is the reconciliation frame error rate. $\chi_{BE}$ is the maximum of the Holevo information that Eve can obtain from the information of Bob. $\bigtriangleup(n)$ is the finite-size offset factor. Because of the imperfect reconciliation scheme, the secret key rate is reduced and the range of the protocol is limited. To achieve a high secret key rate, high reconciliation efficiency $\beta$ is needed under low SNRs, and $\mathrm{FER}$ should be as low as possible.

\section{Design of low-rate RL-LDPC codes}
\label{sec3}
In this section, we show the construction of a low-rate RL-LDPC code. In the information reconciliation step of the CV-QKD system, the MET-LDPC codes are combined with multidimensional reconciliation owing to their capacity-approaching performance, particularly in the low code rate regime. First, we construct a base matrix of the RL-LDPC code with a rate of 0.02 based on the degree distribution of MET-LDPC codes.

Usually, parity-check matrices are used to represent MET-LDPC codes. The rows of matrices represent check nodes, and the columns represent variable nodes. Here, let $\mathbf{d}:=(d_{1},d_{2},...,d_{n_{e}})$ be a multi-edge degree, and let $\mathbf{x}:=(x_{1},x_{2},...,x_{n_{e}})$ denote (vector) variables. Each element $d_{i}$ of $\mathbf{d}$ indicates the number of edges of type $i$ incident to a node. The sum of these elements in $\mathbf{d}$ is referred to as the degree of $\mathbf{d}$. We use $\mathbf{x}^{\mathbf{d}}$ to denote $\prod_{i=1}^{n_{e}}x_{i}^{d_{i}}$. Two multivariate polynomials are used to determine the degree distribution of the MET-LDPC codes; they are as defined
\begin{equation}\label{degree-distribution}
\nu(x):=\sum\nu_{\mathbf{d}}\mathbf{x}^{\mathbf{d}}, ~~~\mathrm{and}~~~ \mu(x):=\sum\mu_{\mathbf{d}}\mathbf{x}^{\mathbf{d}},
\end{equation}
where coefficient $\nu_{\mathbf{d}}$ ($\mu_{\mathbf{d}}$) is the proportion of variable nodes (check nodes) of the multi-edge degree $\mathbf{d}$. In eq.(\ref{degree-distribution}), $\nu(x)$ and $\mu(x)$ are called the degree distributions of variable and check nodes, respectively. The rate of MET-LDPC codes is expressed as
\begin{equation}\label{code-rate}
R=\sum\nu_{\mathbf{d}}-\sum\mu_{\mathbf{d}}.
\end{equation}

For an unpunctured ensemble, the sum of $\nu_{\mathbf{d}}$ is 1 and the sum of $\mu_{\mathbf{d}}$ is $1-R$. Here, the codeword length of the MET-LDPC codes is $N$, and $\nu_{\mathbf{d}}\times N$ ($\mu_{\mathbf{d}}\times N$) is the number of variables (check) nodes with edge type $\mathbf{d}$. The threshold of a code is defined as the maximum standard deviation of the noise channel that can be corrected when the code length is infinite and the maximum number of iterations is sufficiently large. The threshold can be estimated using the density evolution method. Table~\ref{tab:LDPCDIS} presents the degree distributions for the MET-LDPC code rates of 0.05 and 0.02 \cite{wang2017efficient}. The two MET-LDPC codes exhibit excellent error correction performance and are usually used in long-distance CV-QKD systems. The degree distributions possess three edge types, and the multinomial $\nu(x)$ ($\mu(x)$) can be partitioned into three (two) multinomials.

\begin{table}
	\caption{\label{tab:LDPCDIS}The degree distributions of the code rate 0.05 and 0.02 MET-LDPC codes.} 
	\footnotesize\rm
	\begin{ruledtabular}
		\begin{tabular}{c c c}
			Code rate & Degree distribution & Threshold \\
			\hline
			\specialrule{0em}{2pt}{2pt} 
			0.05 & \makecell[l]{$\nu(x) = 0.04x_{1}^{2}x_{2}^{34}+0.03x_{1}^{3}x_{2}^{34}+0.93x_{3}^{1}$\\
				$\mu(x)=0.01x_{1}^{8}+0.01x_{1}^{9}+0.41x_{2}^{2}x_{3}^{1}$\\~~~~~~~~~$+0.52x_{2}^{3}x_{3}^{1}$} & 3.674\\
			\specialrule{0em}{2pt}{2pt}
			0.02& \makecell[l]{$\nu(x)=0.0225x_{1}^{2}x_{2}^{57}+0.0175x_{1}^{3}x_{2}^{57}+0.96x_{3}^{1} $ \\ $\mu(x)=0.010625x_{1}^{3}+0.009375x_{1}^{7}$\\~~~~~~~~~$+0.6x_{2}^{2}x_{3}^{1}+0.36x_{2}^{3}x_{3}^{1}$} &5.91 \\
		\end{tabular}
	\end{ruledtabular}
\end{table}

Next, we consider the MET-LDPC code in Table~\ref{tab:LDPCDIS} as an example to design a RL-LDPC code, as shown in Figure~\ref{fig:raptor-code}. The code length $N$ is $10^{6}$. For convenience, we consider the submatrices $\mathbf{A}$ and $\mathbf{B}$ as one matrix, denoted by $\mathbf{AB}$, and $\mathbf{C}$ and $\mathbf{D}$ as one matrix, denoted by $\mathbf{CD}$. The base matrix of the RL-LDPC code can be obtained by concatenating the submatrices $\mathbf{AB}$, $\mathbf{CD}$, and $\mathbf{I}$. In this MET-LDPC code, there are large numbers of variable nodes with degrees of 1, which are represented by the degree distribution $\nu_{t3}(x)=0.96x_{3}^{1}$ and $\mu_{t3}(x)=0.96x_{3}^{1}$, corresponding exactly to the identity matrix $\mathbf{I}$, with a size of $960000\times960000$. Thus, we only need to design $\mathbf{AB}$ and $\mathbf{CD}$ according to the multinomials. Ref.~\cite{richardson2002multi} shows that the concatenation of a high-rate code and a large number of single parity-check codes can achieve a threshold close to the Shannon limit.
The submatrix $\mathbf{AB}$ comprising edge type-1 can be represented by the degree distribution $\nu_{t1}(x)=0.0225x_{1}^{2}+0.0175x_{1}^{3}$ and $\mu_{t1}(x)=0.010625x_{1}^{3}+0.009375x_{1}^{7}$. This matrix $\mathbf{AB}$ of size $20000\times40000$ can also be expressed using the following polynomials according to its row and column weights:
\begin{align}\label{AB_DEGREE}
\nu_{t1}(x)&=0.5625x^{2}+0.4375x^{3},\nonumber\\
\mu_{t1}(x)&=0.53125x^{3}+0.46875x^{7}.
\end{align}
The above equation shows that there are 22500 columns with a weight of 2, 17500 columns with a weight of 3, 10625 rows with a weight of 3, and 9375 rows with a weight of 7 in matrix $\mathbf{AB}$. Similarly, the submatrix $\mathbf{CD}$ comprising edge type-2 can be represented as
\begin{align}\label{CD_DEGREE}
\nu_{t2}(x)&=x^{57},\nonumber\\
\mu_{t2}(x)&=0.625x^{2}+0.375x^{3}.
\end{align}
\begin{figure}[]
	\centering
	\includegraphics[width=0.9\linewidth]{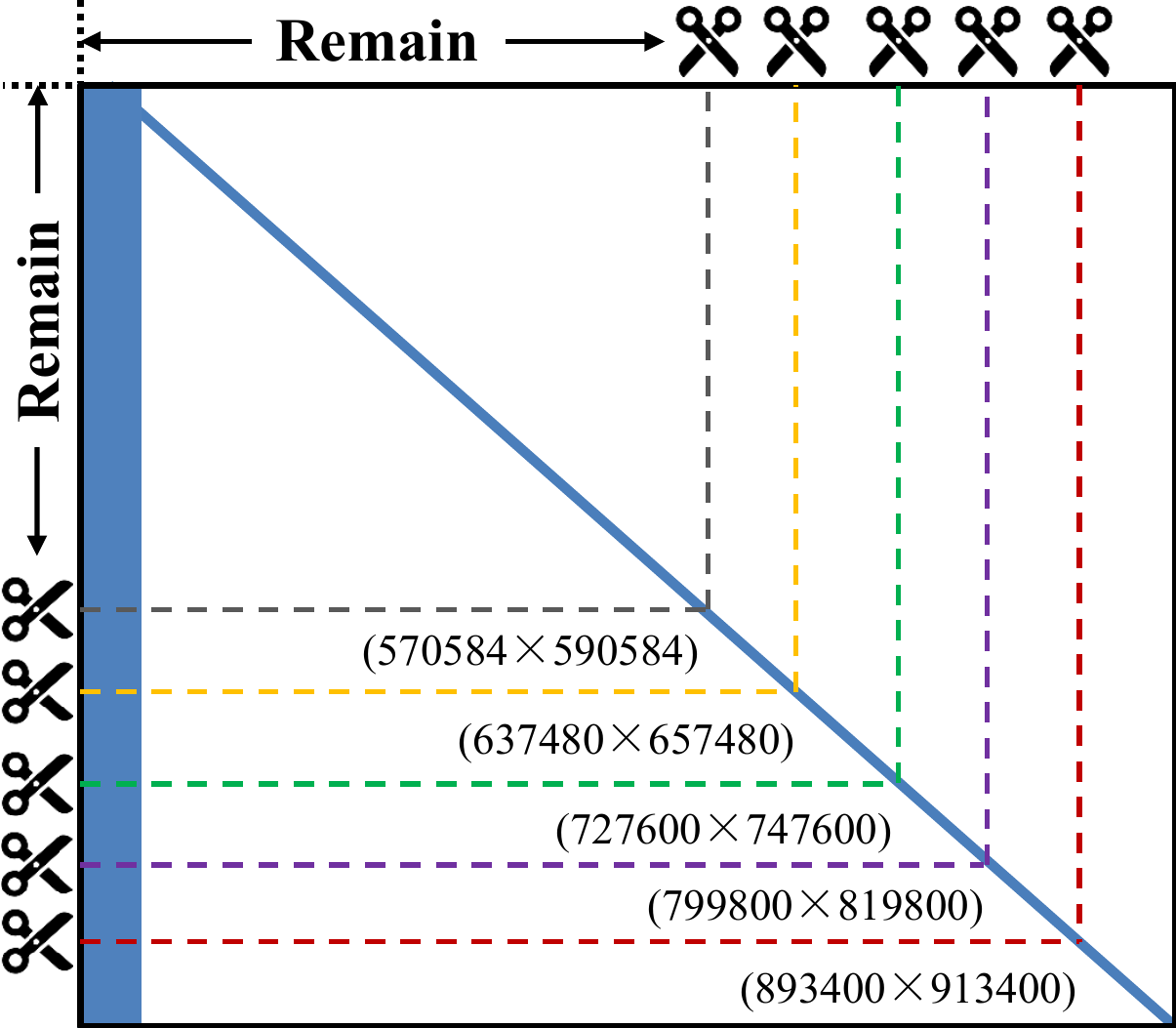}
	\caption{Schematic of cutting the base matrix. The dotted lines indicate the cutting direction and position from the top and left. The upper left corner of the matrix is retained. Different colors represent LDPC codes with different target code rates. The sizes of the five target matrices obtained herein are $893400\times 913400$, $799800\times 819800$, $727600\times 747600$, $637480\times 657480$, and $570584\times 590584$, respectively. The blank area represents the zero elements.}
	\label{fig:cuttingM}
\end{figure}

The random construction method \cite{PEG2005} can be employed to construct the parity-check matrices using the parameters of the above two matrices. Finally, we must concatenate the submatrices simply to achieve the base matrix with a code rate of 0.02. To obtain a high secret key rate in the long-distance CV-QKD system, high reconciliation efficiency under a low SNR is required. Reconciliation efficiency is a significant parameter for evaluating the performance of the information reconciliation step. In the CV-QKD system, reconciliation efficiency can be expressed as
\begin{equation}\label{efficiency_reconciliation}
\beta =\frac{R}{C(s)},
\end{equation}
where $R$ is the code rate, which equals $(N-M)/N$, $C(s)$ is the Shannon capacity, and $C(s)={1}/{2}\mathrm{log}(1+s)$ at SNR $s$ of the additive white Gaussian noise (AWGN) channel. From eq.(\ref*{efficiency_reconciliation}), we know that one code with a fixed rate cannot maintain high reconciliation efficiency under different SNRs.

The introduced RL-LDPC codes exhibit rate compatibility similar to the Raptor codes. Figure~\ref{fig:cuttingM} shows the method for generating different code rates using the designed base matrix. The same length of the base matrix is cut from the right and bottom sides, and the upper left corner of the base matrix is retained. The dotted lines represent the boundary of the target matrices. Assume that the punctured length is $p$; then, the code rate of the target matrix is modified by
\begin{equation}\label{puncture_rate}
R_{p} = \frac{N-p-(M-p)}{N-p} =  \frac{N-M}{N-p}.
\end{equation}
\begin{figure}[]
	\centering
	\includegraphics[width=0.9\linewidth]{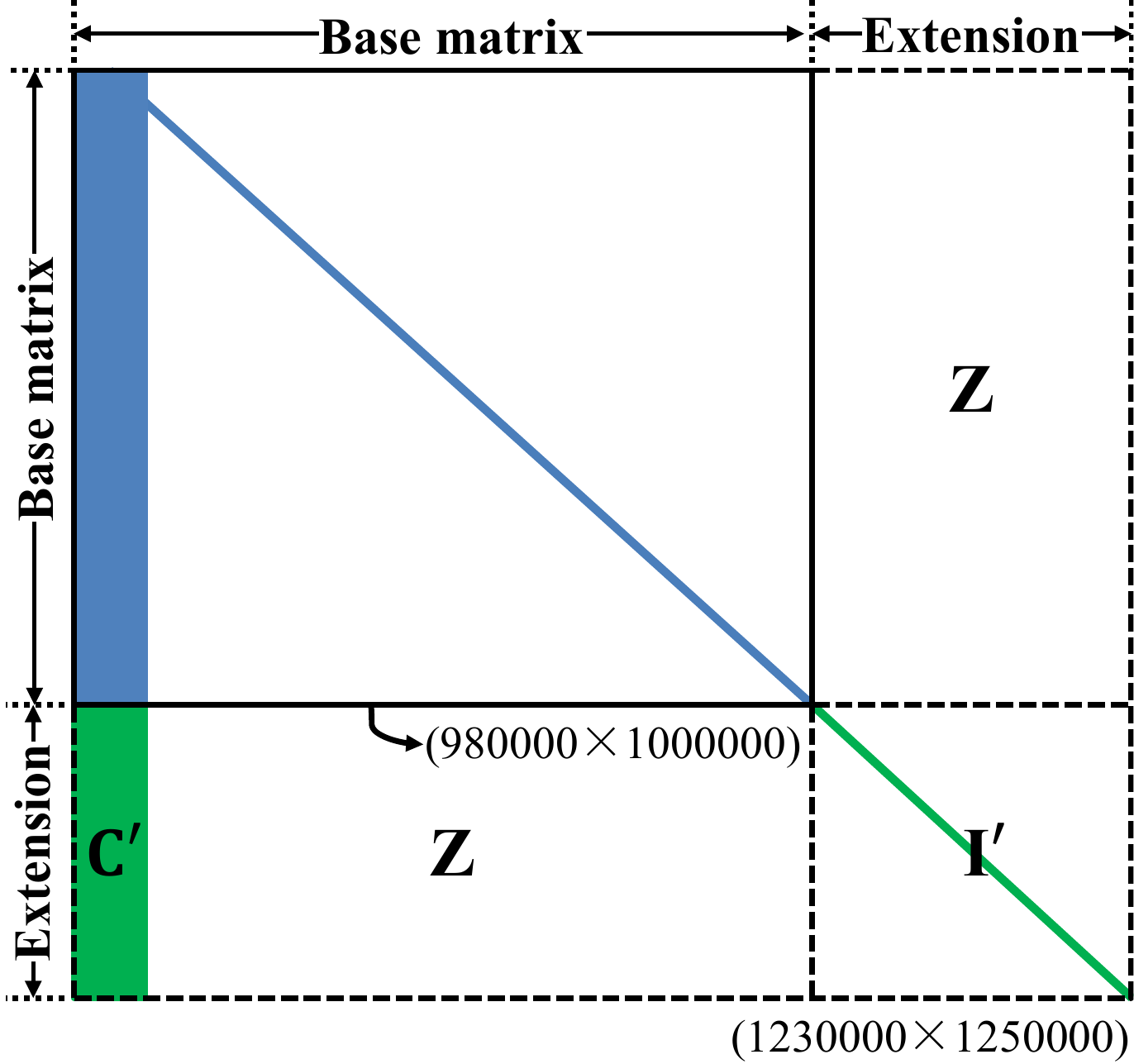}
	\caption{Schematic of extending the base matrix. The solid black line indicates the base matrix with a size of $980000\times 1000000$. The dotted line indicates the part that needs to be extended. The size of the target matrix obtained herein is $1230000\times 1250000$. The blank area and $\mathbf{Z}$ represent the zero elements. $\mathbf{I'}$ represents an identity matrix with a size of $250000\times 250000$. $\mathbf{C'}$ represents the designed extended submatrix.}
	\label{fig:ExternM}
\end{figure}

Although rate compatibility can be easily achieved, the error correction performance of the target matrix is still a challenge. The code rate of the target matrix gradually increases as the base matrix is cut continuously, and its degree distribution also changes. The selection of the degree distribution is an important factor affecting the code performance; thus, the design of the submatrix $\mathbf{CD}$ must satisfy certain rules. In other words, after it is cut, the degree distribution will be changed according to a higher code rate than 0.02. In this work, we selected the code of rate as 0.05 in Table~\ref{tab:LDPCDIS} for reference. Let ${\mu}'_{t2}(x)$ denote the check node degree distribution of submatrix $\mathbf{CD}$ in reference codes. When the code rate of the target matrix gradually increases, ${\mu}'(x)$ must satisfy
\begin{equation}\label{mu_deg_c}
\lim_{R\rightarrow R_{t}}{\mu}'(x)={\mu}'_{t2}(x),
\end{equation}
where ${\mu}'(x)$ denotes the check node degree distribution of submatrix $\mathbf{CD}$ in target codes and $R_{t}$ is the rate of reference codes.

However, it can be seen from eq.(\ref{puncture_rate}) that puncture can only increase the code rate of the designed base matrix. When the SNR reduces, the error correction performance will deteriorate; thus, it is necessary to determine a way to reduce the code rate of the designed base matrix. Fortunately, we can simply reduce the code rate of the RL-LDPC codes and maintain promising performance. Figure~\ref{fig:ExternM} shows the schematic diagram of extending the base matrix. We must only fill in some data on the right and bottom sides of the designed base matrix. More simply, only the extended matrix $\mathbf{C'}$ needs to be designed here. Assume the extended length is $e$; the code rate of the target matrix is modified by
\begin{equation}\label{shorten_rate}
R_{e} = \frac{N+e-(M+e)}{N+e} =  \frac{N-M}{N+e}.
\end{equation}

In theory, ${\mu}'(x)$ will change with an increase in $e$ according to eq.(\ref{mu_deg_c}). A lower rate code than 0.02 should be selected as a reference, and ${\mu}'_{t2}(x)$ denotes its degree distribution. At the time of writing this paper, no degree distribution of code rate less than 0.02 for long-distance CV-QKD systems has been reported and the focus of this paper is not to design it. Thus, we use $\mu_{t2}(x)$ to construct matrix $\mathbf{C'}$. The code rate of the extended matrix is 0.016. Using the above simple steps, we can achieve matrices with code rates between 0.016 and 0.034 based on the designed RL-LDPC code.

\section{Simulation results}
\label{sec4}
\begin{figure}[]
	\centering
	\includegraphics[width=1.0\linewidth]{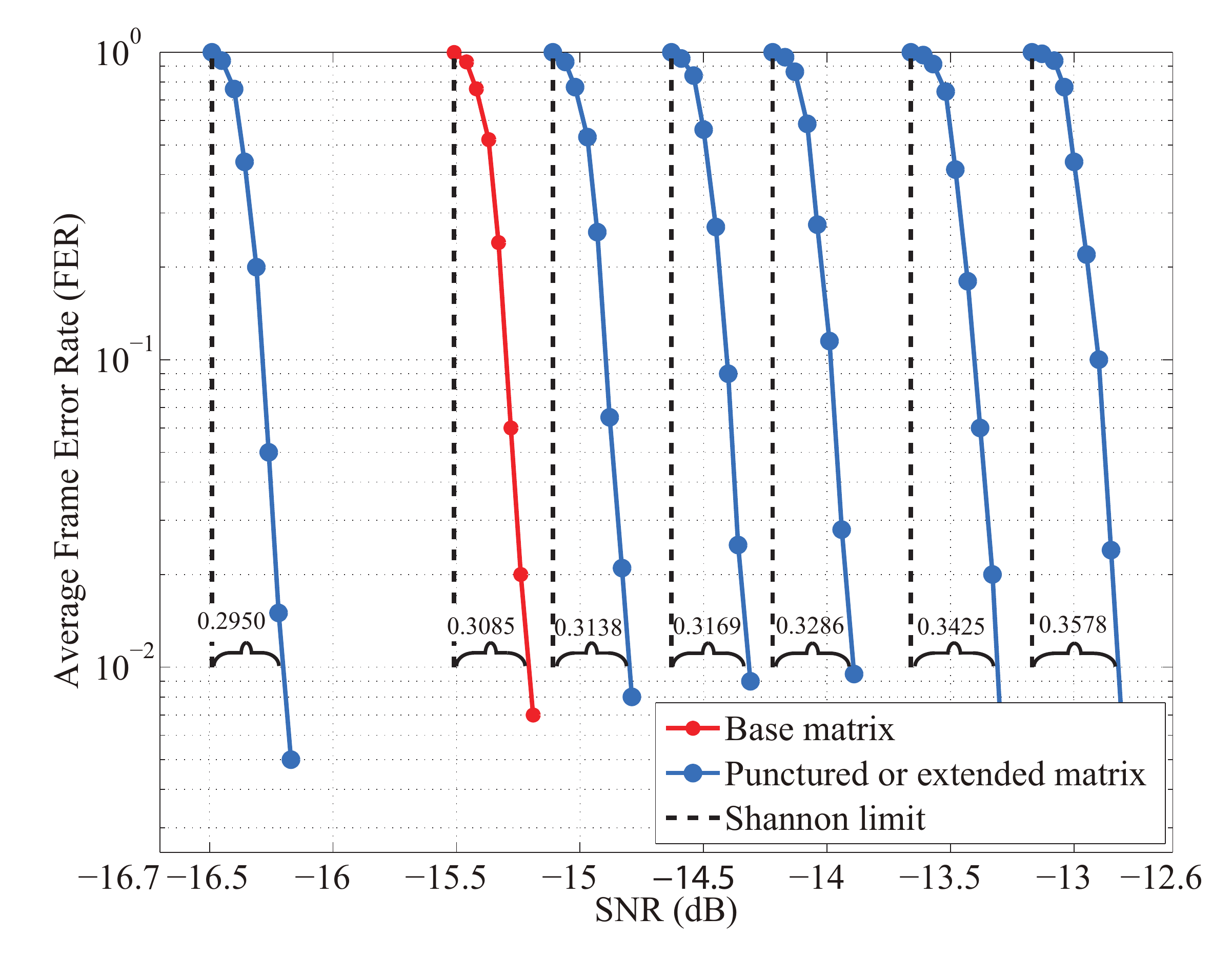}
	\caption{$\mathrm{FER}$ performance of the designed RL-LDPC code with multidimensional reconciliation scheme. The red solid line indicates the designed base matrix of rate 0.02. On its right are punctured submatrices and its left is extended matrix. On each solid line, the solid dots from top to bottom represent the $\mathrm{FER}$ corresponding to the reconciliation efficiency from 100\% to 93\%. The black dotted lines indicate the Shannon limit of each rate. The black braces indicate the gap from Shannon limit of these codes at $\mathrm{FER}=10^{-2}$. The maximum decoding iteration is 400.}
	\label{fig:SNR-FER} 
\end{figure}
In this section, we show the performance of the low rate RL-LDPC code in CV-QKD system. First, we simulate a Gaussian variable with different SNRs in the AWGN channel. Then, the error correction performance is analyzed based on the log-likelihood ratio belief propagation (LLR-BP) using multidimensional reconciliation. Here, we adopt an eight-dimensional reconciliation because it shows the highest performance compared with other dimensions (details for Appendix.~\ref{sec:level1}). Figure~\ref{fig:SNR-FER} shows the $\mathrm{FER}$ of the designed RL-LDPC code. From this figure, we can observe that with increasing $p$, when $\mathrm{FER}=10^{-2}$, the gap from the Shannon limit becomes larger, which implies that the error correction performance is worse. We can conclude that the performance of the smallest two submatrices tends to decrease further because when the size of the matrix becomes smaller, the length of the code becomes shorter as well as the degree distribution of the matrix changes. The error correction performance of the extended matrix with a code rate of 0.016 is slightly better than that of the base matrix, mainly because of the long code length of the former.
\begin{figure}[]
	\centering
	\includegraphics[width=1.0\linewidth]{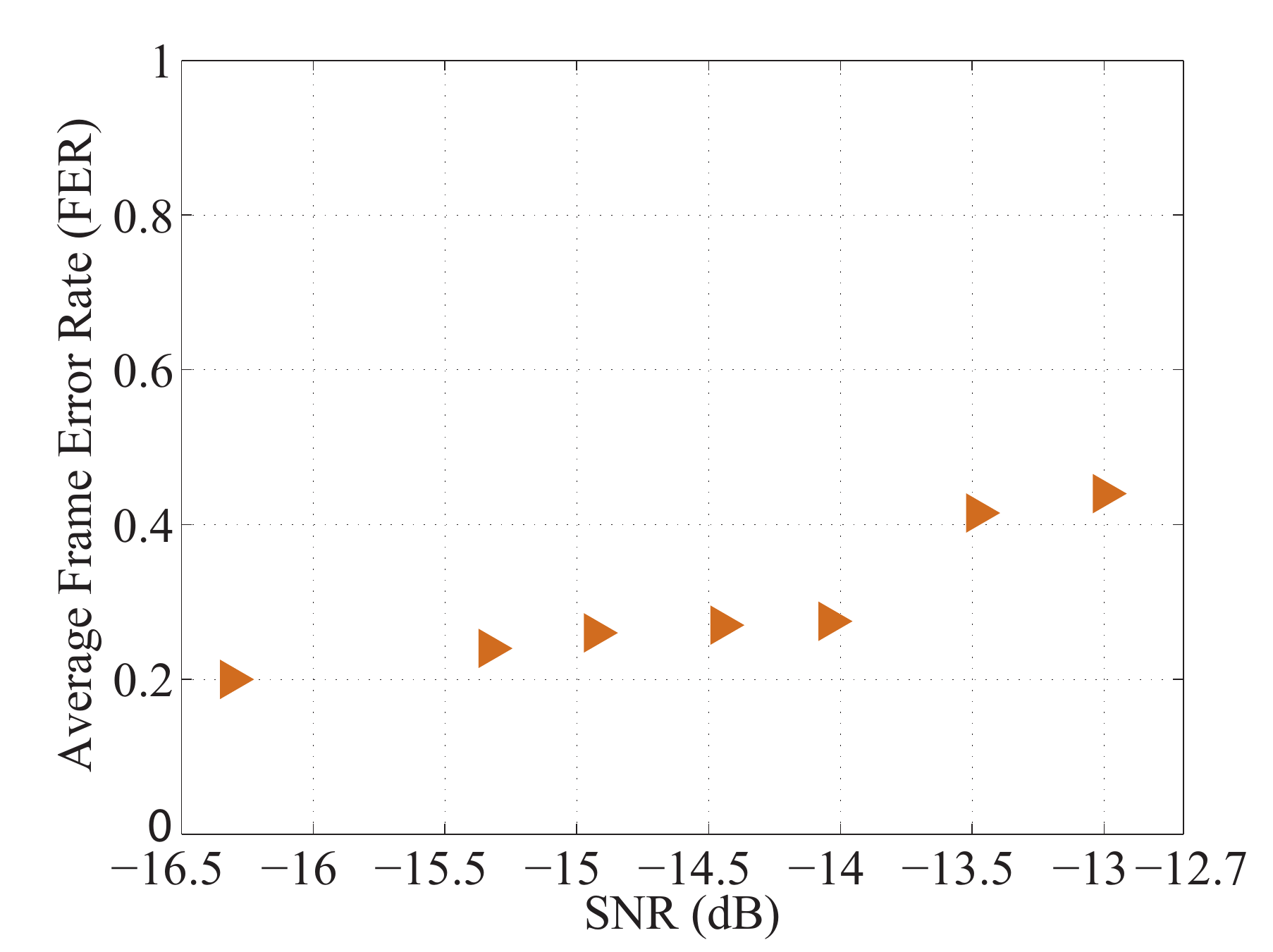}
	\caption{$\mathrm{FER}$ performance of designed RL-LDPC codes with different code rates at reconciliation efficiency $\beta=96\%$ under different SNRs. The code block length and code rate of seven triangles are shown in Figure~\ref{fig:SNR-FER}. Eight-dimensional reconciliation is adopted in this scheme, and the maximum-decoding iteration is 400.}
	\label{fig:SNR_FER2}
\end{figure}
Figure~\ref{fig:SNR_FER2} shows the $\mathrm{FER}$ performance of the RL-LDPC code at reconciliation efficiency $\beta=96\%$. Here, the LLR-BP decoding algorithm is employed using an eight-dimensional reconciliation method. Although the SNR fluctuates significantly, the decoding success rate of the RL-LDPC codes can still be kept stable. These codes can significantly reduce the complexity of constructing a matrix while maintaining the $\mathrm{FER}$ performance.

\newcommand{\tabincell}[2]{\begin{tabular}{@{}#1@{}}#2\end{tabular}}
\begin{table}
	\caption{\label{tab:RLLDPC-SPEED}GPU decoding comparison with different block lengths and code rates in the BP decoder.} 
	\footnotesize\rm
	\begin{ruledtabular}
		\begin{tabular}{c|c c c c p{2cm}|}
			Code rate& 0.016 &0.02&0.02& 0.0304 \\
			\hline
			Block length & $1.25\times10^{6}$ & $10^{6}$ &$10^{6}$ & 657,480\\
			\tabincell{c}{Total number\\ of edges}& 4,187,300 & 3,337,500&3,337,500& 2,181,429\\ 
			\tabincell{c}{Number of\\ codewords}& 32&64&64&96 \\ 
			$\beta$ &98\% &98\% &96\% & 95\%\\
			$FER$ &0.75 &0.76&0.453& 0.375\\
			Max iterations& 400&400&200&200 \\
			\tabincell{c}{Decoding Speed\\ (Mbits/s)}&8.14&8.17&16.41&16.47 \\
		\end{tabular}
	\end{ruledtabular}
\end{table}

The decoding latency of the RL-LDPC code on the GPU platform is tested to meet the real-time processing requirements of the CV-QKD system. We use the parallel decoding method of ref.~\cite{wang2018high00} on a single NVIDIA TITAN Xp GPU. Table~\ref{tab:RLLDPC-SPEED} presents the test results for fore codes. The number of codewords refers to the number of code blocks decoded in parallel using the same block length and code rate. By reducing the maximum iterations, both the decoding speed and $\mathrm{FER}$ can be improved.
The lower the SNR, the higher the required reconciliation efficiency. Therefore, low-rate codes need more iterations to be decoded. In comparison, the maximum iterations of high-rate code can be appropriately reduced to improve the decoding speed.
For a submatrix with a code rate of 0.0304, the total number of edges is one-third less than that of the base matrix. Although more codewords can be processed simultaneously, the decoding speed cannot improve significantly. Thus, the decoding speed is linearly related to the number of iterations when the GPU memory is fully utilized.

\begin{figure}[]
	\centering
	\includegraphics[width=1\linewidth]{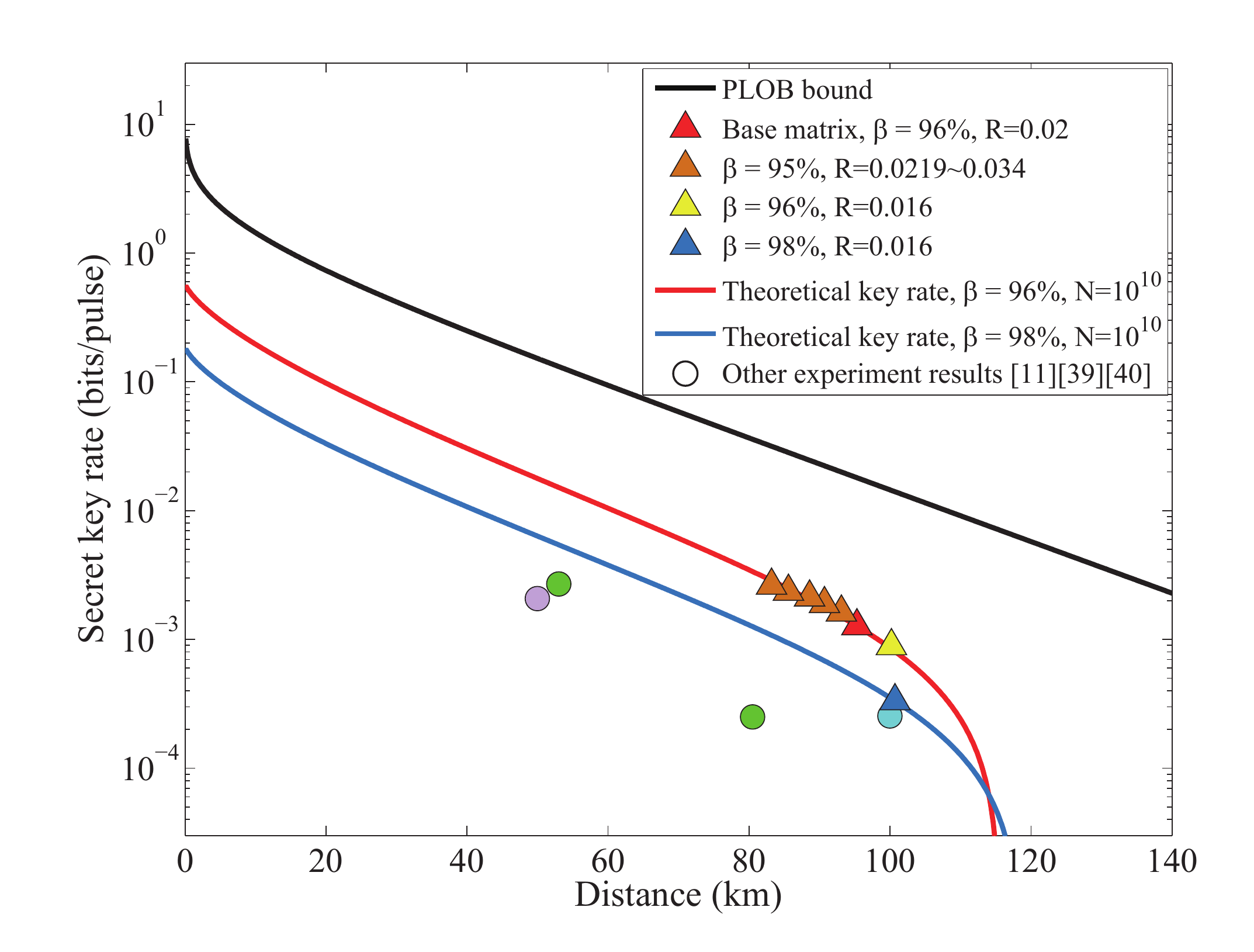}
	\caption{(Color online) Finite-size secret key rate vs. distance. The triangles represent our simulation results. The red triangle represents the base matrix, and the red solid line is the asymptotic theoretical key rates at $\beta = 96\%$. The five orange triangles represent five submatrices after cutting at $\beta = 95\%$. The yellow and blue triangles represent the extended matrix at $\beta = 96\%$ and $\beta = 98\%$, respectively. The blue solid line is the asymptotic theoretical key rates at $\beta = 98\%$. The modulation variance is always maintained at the optimal value, and the above $\mathrm{FER}$ refers to the result in Figure~\ref{fig:SNR-FER}. Other parameters are as follows: excess noise is 0.01, the efficiency of the homodyne detector is 0.6, the standard loss of a single-mode optical fiber cable is 0.2 $dB/km$, and the electric noise is 0.015. For comparison, we also show previous experimental results\cite{jouguet2013experimental,wang201525,huang2016long} and compare the values of our rates with the PLOB bound \cite{pirandola2017fundamental}}.
	\label{fig:skr}
\end{figure}
In Ref.~\cite{zhou2019}, it is shown that the optimal modulation variance can further improve the secret key rate. Our work also allows the system to maintain the optimal modulation variance without sacrificing the target SNR. According to eq.(\ref{finite_secret_key}), the tradeoff between $\mathrm{FER}$ and $\beta$ is considered to obtain high secret key rate.
Furthermore, we can use all the data to estimate the channel parameter and extract the secret keys \cite{Wang2019}.

Figure~\ref{fig:skr} shows the finite-size secret key rate of the designed RL-LDPC codes with the transmission distance. The eight triangles are the simulation results of seven designed matrices under different $\beta$. Here, we consider that the selection of $\mathrm{FER}$ and $\beta$ will maximize the results, and the block length is $10^{10}$. The yellow and blue triangles represent the extended matrix with a code rate of 0.016 at $\beta = 96\%$ and $\beta = 98\%$, respectively.
Notably, the secret key rate can be extracted at high speed when the SNR is as low as 0.0226 (-16.45 dB). Although a high reconciliation efficiency is suitable for long distances, the low $\mathrm{FER}$ can significantly improve the secret key rate. The system repetition rate is not considered in this simulation result. Other experimental results \cite{jouguet2013experimental,wang201525,huang2016long} with the standard loss of a single-mode optical cable are also shown in Figure~\ref{fig:skr}. Here, we also compared the key rate with the PLOB bound \cite{pirandola2017fundamental}, i.e., the fundamental limit of repeaterless quantum communications. The RL-LDPC codes offer the advantages of practical applications in the CV-QKD system.

\section{Conclusion}
\label{sec5}
The practical application of the CV-QKD system is increasingly urgent and requires high -performance postprocessing, particularly in terms of reconciliation efficiency, frame error rate, and processing speed. Error correction steps are the key to restrict the above factors. This paper introduces RL-LDPC codes into the CV-QKD system. These codes exhibit the rateless property of Raptor codes as well as good error correction performance of MET-LDPC codes under a low SNR. We design the RL-LDPC matrix with a code rate of 0.02 and randomly select six target rates as reference. The seven designed matrices show positive performance and can meet the system requirements. Furthermore, our work allows the system to operate under the optimal modulation variance to improve the secret key rate. Our work can promote the application of the CV-QKD system in a practical environment.

In our work, we do not consider designing a matrix in the lower-triangular form to reduce encoding complexity because the improvement in encoding is not significant on the GPU platform. If the application is considered in a small integrated system, such as a field-programmable gate array platform, the lower-triangular form will be a good choice. Using quasi-cyclic codes to construct the RL-LDPC matrix can reduce memory usage and decoder implementation complexity.

\section{Acknowledgments}
This work was supported by the Key Program of National Natural Science Foundation of China (Grant No. 61531003), the National Natural Science Foundation of China (Grant No. 62001041), and the Fund of State Key Laboratory of Information Photonics and Optical Communications.

\begin{appendix}
\label{11111}
	

	


	\section{}

	\subsection{\label{sec:level1} Multidimensional reconcliaition method}
	The central idea of multidimensional reconciliation is to reformulate the attenuated Gaussian physical channel into a virtual Binary input AWGN (BI-AWGN) channel. Here, we consider an eight-dimensional reconciliation as an example to introduce the multidimensional reverse reconciliation method. After the quantum transmission and sifting steps, Alice and Bob share correlated Gaussian sequences. The detailed analysis of a set of data is presented. First, they combine eight continuous-variable sequences into an eight-dimensional vector. Let $X=(x_{1},x_{2},...,x_{8})^{T}$ and $Y=(y_{1},y_{2},...,y_{8})^{T}$ corresponding to the correlated Gaussian vectors of Alice and Bob, respectively; then, $X=Y+Z'$ with $Z'\sim (0,\sigma'^{2})$, where $\sigma'^{2}$ is multidimensional noise variance. Then, Alice and Bob normalize their Gaussian variables $X$ and $Y$ to $x=\frac{X}{\left \| X \right \|}$ and $y=\frac{Y}{\left \| Y \right \|}$, respectively, where $\left \| X \right \|=\sqrt{\left \langle X,X \right \rangle}=\sqrt{\sum_{i=1}^{8}x_{i}^{2}}$ and $\left \| Y \right \|=\sqrt{\left \langle Y,Y \right \rangle}=\sqrt{\sum_{i=1}^{8}y_{i}^{2}}$. The vectors $x$ and $y$ are uniformly distributed on the unit sphere $\mathbb{S}^{7}$ of $\mathbb{R}^{8}$. Bob randomly generates a binary sequence $(b_{1},b_{2},...,b_{8})$, and maps it to the unit sphere:
	\begin{equation}\label{app1}
	u=(\frac{(-1)^{b_{1}}}{\sqrt{8}},\frac{(-1)^{b_{2}}}{\sqrt{8}},...,\frac{(-1)^{b_{8}}}{\sqrt{8}}).
	\end{equation}
	
	Next, Bob calculates the mapping function $M(y,u)$ with $M(y,u)x=u$ using the following formula:
	\begin{equation}\label{app2}
	M(y,u)=\sum_{i=1}^{8}\alpha _{i}(y,u)A_{i},
	\end{equation}
	where $\alpha (y,u)=(\alpha _{1},\alpha _{2},...,\alpha _{8})^{T}$ is the coordinate of the vector $u$ under the orthonormal basis $(A_{1}y,A_{2}y,...,A_{8}y)$ and it can be expressed as $\alpha (y,u)=(A_{1}y,A_{2}y,...,A_{8}y)^{T}u$. $A_{i}(i=1,2,...,8)$ is the orthogonal matrix of size $8\times 8$ and has been provided in the Appendix of Ref.~\cite{leverrier2008multidimensional}.
	
	Then, Bob transmits the function $M(y,u)$ to Alice through a public classical authenticated channel. Alice operates the same data mapping on x to map the Gaussian variables to $v=M(y,u)x$, which is the noisy version of $u$.
	
	After the above steps, the virtual BI-AWGN channel with binary input $u$ and continuous -variable output $v$ is established. Then, Alice uses the RL-LDPC codes to recover $u$ from $v$.

\end{appendix}

\end{document}